\documentstyle{aipproc}

\begin{document}
\title{Unscrambling New Models for Higher-Energy Physics
\thanks{Plenary talk at the 5th International Linear Collider Workshop (LCWS 2000),
Fermilab, October 24-28, 2000.}}

\author{Bogdan A. Dobrescu \thanks{This work 
was supported by DOE under contract DE-FG02-92ER-40704.}}
\address{Department of Physics, Yale University\\
New Haven, CT 06511, USA}

\maketitle

   \vspace*{-6.4cm}
\noindent \makebox[12.3cm][l]{\small \hspace*{-.2cm}
hep-ph/0103038} {\small YCTP-P1-02 } \\
\makebox[10.8cm][l]{\small \hspace*{-.2cm} February 28, 2001 }
\\

 \vspace*{4.7cm}

\begin{abstract}
There is strong evidence that new physical degrees of freedom and new
phenomena exist and may be revealed in future collider experiments.
The best hints of what this new physics might be are provided
by electroweak symmetry breaking. I briefly review certain theories 
for physics beyond the standard model, including the top-quark seesaw 
model and universal extra dimensions. A common feature of these models
is the presence of vector-like quarks at the TeV scale.
Then I discuss the role of a linear $e^+e^-$ collider in 
disentangling this new physics.
\end{abstract}

\baselineskip=18pt \pagestyle{plain} \setcounter{page}{1}

\section*{The case for new physics}
The observed fundamental particles, namely the 
$SU(3)_C \times SU(2)_W \times U(1)_Y$ gauge bosons, the
longitudinal degrees of freedom of the $W^\pm$ and $Z^0$, and 
three generations of quarks and leptons, may explain in principle 
all observed physical phenomena. However, there is strong 
evidence for the existence of new
phenomena at higher energy scales than the ones probed so far. 
The most robust argument is provided by 
the perturbative violation of unitarity in the $WW$ scattering,
at a scale of order 1~TeV \cite{Lee:1977yc}.
Therefore, either the $W$ and $Z$ have strongly coupled
self-interactions at the TeV scale, or new fundamental
degrees of freedom exist.
The scale of these new phenomena is within the reach of 
future collider experiments, and exploring them is at the heart
of high-energy physics.

The standard model accomodates rather well all
available data, especially when the Higgs boson is light
\cite{ewwg}.
More generally, any model with a decoupling limit \cite{Appelquist:1975tg}
given by the standard model is viable, at least in that limit. 
These are models in which all the particles beyond the standard model 
may be given large electroweak-symmetric masses.
I will refer to them as ``decoupling models''.
In practice, the decoupling may be only partial,
so that new particles with electroweak-symmetric 
masses of order 1 TeV give rise at the electroweak scale 
to higher-dimensional operators which may change the fit of the 
standard model to the data. In particular, the Higgs boson mass
may be larger in this case, even close to the triviality bound 
\cite{Chivukula:2000px}.
In what follows I will concentrate on decoupling models, 
but one should keep in mind that it is not currently 
possible to rule out all models without a Higgs boson, because 
the physical quantities relevant for comparing with the data
typically cannot be computed when the interactions are non-perturbative. 


Even if a standard-model-like Higgs boson will be discovered, and maybe 
for a while no experimental data will hint at other new physics, there are
robust theoretical reasons to expect physics beyond the standard model.
First, the Higgs self-coupling increases with the energy and the theory 
breaks down at some scale.
Second, the ${\rm U}(1)_Y$ gauge coupling is also ill-behaved at high 
energy.
Third, the standard model does not include gravity, while the 
measured gravitational effects are produced by matter composed of 
standard model particles. Additional reasons to expect new physics 
are given by the large number of parameters 
in the standard model, the large hierarchies between their values, 
and the rather strange field content of the standard model
(for example, the Higgs doublet is the only scalar field, and 
does not appear to be {\it theoretically} motivated.)

It would be highly desirable that 
a theory which includes both the standard model and gravity 
explains why the electroweak interactions
are so much stronger than the gravitational interactions.
Currently, the only possible explanations for this hierarchy between
the electroweak and Planck scales require new particles with mass
of order 1 TeV.
The decoupling models may be classified according 
to the solution for the hierarchy problem into
supersymmetric extensions of the standard model, theories with 
a composite Higgs doublet, and theories with extra dimensions.
Although convenient, this classification is not clear-cut given that 
there are supersymmetric models in extra dimensions 
\cite{Antoniadis:1990ew,Dienes:1999vg},  
models of extra dimensions that generate composite Higgs doublets
\cite{Dobrescu:1999dg,Cheng:2000bg,Arkani-Hamed:2000hv}, as well as
 supersymmetric models of Higgs compositeness \cite{Luty:2000fj}.  

The supersymmetric extensions of the standard model and theories 
with extra dimensions accessible only to gravity have been covered
in other talks at this conference \cite{godbole}, so here I review only
models with composite Higgs doublets and/or standard model fields in 
extra dimensions.
I also discuss their implications for experiments at a linear $e^+e^-$ 
collider.


\section*{Top quark seesaw-theory}


The quarks and leptons are chiral, in the sense that their
left- and right-handed components have different $SU(2)_W \times U(1)_Y$
charges. New chiral fermions would have large contributions to the 
electroweak observables and the current data place strong constraints 
on their number, charges and mass splittings. On the other hand, 
vector-like fermions, defined as having the same gauge charges for
the left- and right-handed components, are much less constrained.
This can be seen from the fact that their masses preserve the electroweak 
symmetry, so that in the limit where their mass $m_v$ is much above the 
electroweak scale $v \approx 246$ GeV, their contribution to the 
electroweak observables is suppressed by $v^2/m_v^2$ \cite{Popovic:2000dx}.  

If a vector-like fermion has the same $SU(3)_C \times SU(2)_W \times U(1)_Y$
charges as a standard model fermion, then it may couple to the Higgs doublet.
In the case where this coupling is large, or the number of vector-like 
fermions is large, the properties of the Higgs field change rapidly at scales
above the fermion masses.
To see this, consider a vector-like quark $\chi$ that has the same charges as the
right-handed top quark $t_R$.
Then the most general renormalizable coupling (up to an $U(2)$ global transformation 
acting on $t_R, \chi_R$)
of $\chi$ to the Higgs doublet $H$
may be written as
\begin{equation}
{\cal L} = - \left( \overline{q}_L^3 \ , \ \overline{\chi}_L  \right)
\left( \begin{array}{rl} 0 \;\;\;  & \xi_\chi H \\
        m_{\chi t} & m_{\chi \chi} \end{array} \right)
\left( \begin{array}{c} t_R \\ \chi_R \end{array} \right) + {\rm h.c.}
\end{equation}
Here $q_L^3$ is the top-bottom left-handed doublet,
$\xi_\chi$ is a Yukawa coupling constant, while $m_{\chi t}$ and
$m_{\chi \chi}$ are electroweak symmetric masses.
The eigenvalues of the mass matrix obtained by replacing the Higgs doublet 
with its VEV represent the top quark mass, $m_t$, and the mass $m_\chi$ of 
the new quark:
\begin{equation}
m_{t, \chi}^2 = \frac{1}{2} 
\left( m_{\chi \chi}^2 + m_{\chi t}^2 + \xi_\chi^2\frac{v^2}{2}  \right)
\left[1 \mp \sqrt{ 1 - 2 \left(\frac{\xi_\chi v \, m_{\chi t}}
{m_{\chi \chi}^2 + m_{\chi t}^2 + \xi_\chi^2 v^2/2} \right)^{\! 2}\, } \, 
\right] ~. 
\label{seesaw}
\end{equation}

If the Yukawa coupling is large at the scale $m_\chi$, then 
its scale dependence is very strong and it blows up  
at a scale $\Lambda_c$ not far above $m_\chi$. The dominant contribution 
to this effect is the Higgs wave function renormalization shown in   
Fig.~1. In fact this would be the only leading contribution if the number of 
colors $N_c$ were very large (rather than $N_c =3$), which may be seen 
as an indication that this phenomenon persists at strong coupling.

\vspace*{3.5cm}
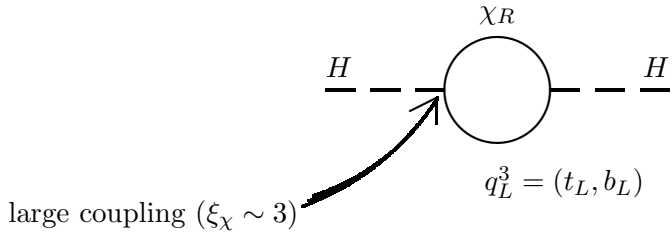
\begin{figure}[h!] 
\centerline{\hspace*{-6.9cm}
\begin{picture}(388,40)(-195,-50)
\thicklines
\put(96,50){ \circle{40}}
\put(80,50){\line(-2, 0){11}}
\put(63,50){\line(-2, 0){11}}
\put(46,50){\line(-2, 0){11}}
\put(120,50){\line(2, 0){11}}
\put(137,50){\line(2, 0){11}}
\put(154,50){\line(2, 0){11}} 
\put(95,12){\small $q^3_L=(t_L,b_L)$}
\put(35,55){\small $H$}
\put(93,77){\small $\chi_R$}
\put(155,55){\small $H$}
\qbezier(77,47)(60,20)(30,10)
\qbezier(77,47)(59,19)(28,8)
\qbezier(77,47)(58,18)(26,6)
\put(77,47){\line(0, -1){10}}
\put(77,47){\line(-2, -1){10}}
\put(-85,0){\small large coupling ($\xi_\chi \sim 3$)}
\end{picture}
}
\vspace{-.8cm}
\caption{Vector-like quark contribution to the Higgs self-energy.}
\label{fig1}
\end{figure}
\vspace*{.3cm}

The Yukawa coupling that seemingly blows up is a signal for new physics at
the scale $\Lambda_c$. The simplest interpretation is that the scalar $H$
ceases to be a physical degree of freedom above  $\Lambda_c$. A confirmation for 
this interpretation is given by viewing the one-loop graph in Fig.~1 as a 
correction to the Higgs kinetic term rather than a change in the Yukawa coupling.
Normalizing the coefficient of the Higgs kinetic term to unity at the electroweak 
scale, one finds that it vanishes at the scale $\Lambda_c$ because the correction 
is negative and sufficiently large at that scale. Therefore, the Higgs doublet is
no longer a propagating field at higher scales.
It is then natural to assume that $H$ is a composite field, with some constituents
bound by a strongly-coupled interaction. Furthermore, the large Yukawa coupling 
reveals the constituents: $\chi_R$ and $\overline{q}_L^3$. This implies that
the strongly-coupled interaction should not confine, otherwise the top-bottom
left-handed doublet would not be present in the effective low-energy theory.
$\Lambda_c$ is  called the compositeness scale, and may be in the
TeV range if the Yukawa coupling is sufficiently large at the electroweak scale
($\xi_\chi \sim 3-4$ based on the large-$N_c$ computation.)
The exchange of the non-propagating $H$ field at the compositeness scale
induces a four-quark contact term of the form 
$(\overline{q}_L^3 \chi_R)(\overline{\chi}_R q_L^3)/\Lambda_c^2$ with a 
large dimensionless coupling. This interaction 
represents the dominant effect of the 
strongly-coupled interaction responsible for binding the Higgs doublet,
and is indeed non-confining, being weak at long-distance.

The model of Higgs compositeness described so far is called the top-quark seesaw
theory \cite{Dobrescu:1998nm,Chivukula:1999wd}, because in the limit
$\xi_\chi v, m_{\chi t} \ll m_{\chi \chi}$ the top mass is suppressed 
compared with the electroweak asymmetric fermion mass, $\xi_\chi v/\sqrt{2}$, 
by a seesaw mechanism. 

The Higgs self-coupling, $(-\lambda_h/2) (H^\dagger H)^2$, is also affected by 
the large Yukawa coupling. The one-loop RGE for $\lambda_h$ is given by
\begin{equation}
\frac{d \lambda_h}{d \ln\mu} = \frac{3}{4 \pi^2}\left(\lambda_h \xi_\chi^2 
- \xi_\chi^4 + \lambda_h^2 \right) ~.
\end{equation}
The Higgs boson mass is determined by the self-coupling, $M_h = \sqrt{\lambda_h} v$.
The case where the Higgs boson is light, $\lambda_h \ll \xi_\chi^2$,
implies that $\lambda_h$ decreases at higher scales. 
This situation would be hard to reconcile with the fact that the
Higgs self-coupling
is given by the strongly-coupled interaction between the Higgs 
constituents. One may infer that $\lambda_h$ is rather large and positive 
at the electroweak scale, so that the Higgs boson is heavy, with a
mass $M_h$ estimated to be of order 500 GeV 
\cite{Chivukula:2000px,private}. 
This Higgs mass prediction has clearly large theoretical uncertainties 
due to the non-perturbative 
interactions involved in compositeness, but the large Higgs self-coupling
appears to be a rather generic consequence of the strongly-coupled binding
 interaction.
It is possible though that there exist  several composite Higgs  
fields \cite{Chivukula:1999wd}, in which case the mixing among the various 
CP-even Higgs bosons may allow the lightest one to have standard-model-like
couplings and a mass close to the current LEP II bound \cite{Dobrescu:2001gv}.  

Whether the Higgs boson is as light as $\sim 115$ GeV, or as heavy as
500 GeV, the top-quark seesaw theory remains viable: in the former case
it would be in the decoupling limit $m_\chi \gg v$;
in the latter case it would require the vector-like quark to have
a mass closer to the TeV scale, such that its isospin-violating effects
render the electroweak observables in agreement with the current data 
\cite{Chivukula:2000px,Collins:2000rz}.




When the effective theory
below the compositeness scale includes an extended Higgs sector,
it is possible that a light Higgs boson, with nearly standard 
couplings to fermions and gauge bosons, has completely non-standard
decay modes. This happens whenever
a $CP$-odd scalar has a mass less than half the Higgs
mass and the coupling of the Higgs to a pair of $CP$-odd scalars is not
suppressed. The Higgs boson decays into a pair of $CP$-odd
scalars, each of them subsequently decaying into a pair of
standard model particles, with model dependent branching
ratios~\cite{Dobrescu:2000jt}.
A linear $e^+ e^-$ collider may prove very useful in disentangling the
composite nature of the Higgs boson with non-standard decays, 
by measuring its width and branching ratios.

The heavy quark constituent of the Higgs has a mass $m_\chi$ 
of a few TeV. Nevertheless, a linear $e^+ e^-$ collider operating at
$\sqrt{s} = 500$ GeV, or above, may already see the effects of the 
mixing between the top and $\chi$ \cite{Popovic:2001cj}. For this it will be
necessary to determine the $Z\bar{t}t$ coupling at the few percent level.

The interaction responsible for binding the Higgs field, 
and approximated with 
a four-quark interaction at the $\Lambda_c$ scale, is
provided by a spontaneously broken gauge symmetry, such
as topcolor~\cite{Hill:1991at}, or some flavor or family
symmetry~\cite{Burdman:1999vw}.
Such interactions are  non-confining, and also asymptotically
free, allowing for a solution to the hierarchy problem. 
Above the compositeness scale there must be some additional 
physics that leads to the spontaneous breaking of the
gauge symmetry responsible for binding the Higgs. This may involve 
new gauge dynamics~\cite{Collins:2000rz}, or fundamental scalars and
supersymmetry.
A linear $e^+ e^-$ collider could  study these
interesting strongly-interacting particles only if it operates at 
energies above a TeV. Evidently, the CLIC design, with the acceleration 
provided by a high-intensity beam that allows a center of mass energy 
of a few TeV, is highly desirable in this context.

Other models of Higgs compositeness have been proposed
recently~\cite{Georgi:2001wt,He:2000vp,Aranda:2000vk}, and 
each of them has interesting phenomenological consequence with implications
for a  linear $e^+ e^-$ collider.

\section*{Standard model in extra dimensions}

Extra spatial dimensions accessible to
standard model particles are constrained by the Tevatron and LEP
data to be smaller than of order $(1~{\rm TeV})^{-1}$.
The existence of TeV-size extra dimensions is a logical
possibility, motivated by various theoretical considerations,
such as the generation of hierarchical quark and lepton masses, or
the potential for gauge coupling unification at a scale in the TeV
range~\cite{Dienes:1999vg}.


\subsection*{Kaluza-Klein resonances}

The immediate consequence of this scenario is the existence
of towers of Kaluza-Klein (KK) modes for the particles that
propagate in the TeV-size extra dimensions.
For example, the gluons would have spin-1 color-octet excitations. 
Their masses are given by $\sqrt{j}/R$ where $R$ is the radius of
the extra dimensions and $j$ is an integer that labels the
KK level. The number of states on each level is a function
of the number of extra dimensions.
For example, with one extra dimension, the only occupied levels have
$j = k^2$ where $k$ is an integer. 
Both the density of occupied levels
and the average number of states ($D_n$) on a level increase with the
number of extra dimensions.

The $W, Z$ and photon would have color-singlet
spin-1 excitations with a mass spectrum similar to the KK
gluons, slightly perturbed due to the electroweak symmetry breaking.
In the often considered case where the quarks and leptons 
are localized on a three-dimensional domain wall (a 3-brane), the KK excitations 
of the gauge bosons
have the same couplings up to a factor of order one as the corresponding
standard model states.     
Therefore, the $Z$ and photon KK states may be produced in the $s$ channel
in $e^+e^-$ collisions.
At the same time, the KK excitations of the electroweak gauge bosons 
contribute at tree-level
to the electroweak observables, and are constrained to lie above $\sim 4$ TeV 
in the absence of other compensating effects.
If some standard model fermions propagate in extra dimensions,
for each of these chiral quarks and leptons there is  
a tower of vector-like fermions with mass separations of order $1/R$.
A  linear $e^+e^-$ collider with $\sqrt{s} = 0.5$--1~TeV is unlikely
to produce directly any of these KK excitations, but is very
sensitive to their presence via virtual effects.
%
%
With $\sqrt{s}$ in the TeV range, however, a linear $e^+e^-$ collider  
may produce a series of KK resonances,
which not only would establish the existence of extra dimensions but also
 would  determine the number of extra dimensions
and their structure.

A qualitatively distinct case is that {\it all} 
standard model particles propagate in extra dimensions.
These are called universal extra dimensions.
The KK number is then conserved at each vertex, 
so that the KK excitations may be produced only in groups 
of two or more. Hence, the direct bounds from the Tevatron 
and LEP are significantly lower. Moreover, the KK states
do not contribute at tree level to the electroweak observables. 
The current mass bounds on the first KK states are as low as 300 GeV
for $\delta = 1$ universal extra dimension, and in the 
$400-800$ GeV range for $\delta = 2$ \cite{Appelquist:2000nn}.
These loose bounds make the universal extra dimensions 
 particularly interesting for
collider experiments. At $\sqrt{s} > 600$ GeV, a linear $e^+e^-$ collider    
could already  pair-produce KK leptons, quarks and 
gauge bosons. One possibility is that the KK states decay 
outside the detector, so that the signal would be pairs of highly 
ionizing tracks. An alternative is that some interactions that do not
conserve momentum in the extra dimensions allow the KK states 
to decay promptly into pairs of 
standard model particles.

\subsection*{Composite Higgs from universal extra dimensions}

Apart from these phenomenological implications, phenomena
due to the extra dimensions shed a new light on the origin of 
electroweak symmetry breaking.
An interesting example is provided by the universal extra dimensions.
The KK excitations of the standard model gauge
bosons and fermions give rise to a scalar bound state with 
the quantum numbers of the standard model Higgs
doublet~\cite{Arkani-Hamed:2000hv}. 
The Higgs boson appears as a composite scalar with 
a combination of KK modes of the top-quark playing the
role of constituents. 
This can be easily understood given that the KK excitations
of the gluons and electroweak gauge bosons induce a strong
attractive interaction between the left- and right-handed top-quark
fields. 
It is remarkable that out of the many possible bound states involving
the quarks and leptons, the most deeply-bound state
has the quantum numbers of the Higgs doublet.
This state acquires a vacuum expectation value and, thus,
breaks the electroweak symmetry.
Furthermore, this composite Higgs doublet has a large Yukawa coupling
to the top-quark. Hence, electroweak symmetry breaking and the
 large top mass 
are direct consequences of the experimentally determined gauge
charges of the quarks and leptons.


At a scale in the TeV range, called for convenience the string scale,
$\Lambda_s$, the only degrees of freedom 
are the $SU(3)_C \times SU(2)_W \times U(1)_Y$ gauge bosons and the
three generations of quarks and leptons, all of them propagating 
in a higher-dimensional spacetime.
(At even higher energy scales, the fundamental degrees of freedom of a
theory incorporating quantum gravity are expected to become relevant,
such as the winding modes of string theory.)
Below the scale $\Lambda_s$, fermion--anti-fermion pairs bind via
the $SU(3)_C \times SU(2)_W \times U(1)_Y$ interactions.
Then, below the scale $1/R < \Lambda_s$ that sets the
size of the extra dimensions, the physics is 
described by an effective four-dimensional theory.
This effective theory is the usual standard model,
with the possible addition of a few
other scalars, such as the heavy states of a two-Higgs-doublet model.

This simple model is consistent with all the experimental
data, and is also predictive. The top mass is predicted 
with an uncertainty of about 20\% and is consistent with the experimental
value. This and, more importantly, the 
prediction of the Higgs quantum numbers are unmatched by the 
standard model or its supersymmetric extensions.
However, like the standard model or the MSSM, this model only 
accommodates without predictions the light quark and lepton masses.

The Higgs mass also is determined theoretically,
by solving the RGE for
$x_H \equiv \lambda_h/\lambda_t^2$, where $\lambda_h$ is the 
Higgs boson self-coupling and $\lambda_t$ is the top Yukawa coupling 
($\lambda_t \approx 1$ at the electroweak scale). Ignoring the 
electroweak gauge couplings, the one-loop RGE for $x_H$ is given in 
the case of two  universal extra dimensions by 
\begin{equation}
\frac{d \ln x_H}{d \ln \mu} = \frac{\lambda_t^2(\mu) N_{\rm KK}(\mu)}{(4\pi)^2}
\left[ 12 x_H + 9 - \frac{24}{x_H} + \frac{64 g_3^2(\mu)}{3 \lambda_t^2(\mu)} \right]
~.
\end{equation}
Due to the strong scale-dependence of the number of KK modes lighter than 
the scale $\mu$, $N_{\rm KK}(\mu)$,
the solution of this RGE lies in a narrow interval, being rather 
insensitive to the value of $x_H$ at $\Lambda_s$. The Higgs  
boson mass is predicted in turn to lie
between the $WW$ threshold and the electroweak scale \cite{Arkani-Hamed:2000hv}.
The main theoretical uncertainty in this computation comes from the 
possible existence of other composite scalars, such as the heavy states of a 
two-Higgs-doublet sector. The mixing between the CP-even states could 
in principle reduce the mass of the lightest Higgs boson.

Assuming that the effects of heavy scalars may be ignored, 
the Higgs mass prediction implies that the Higgs boson decays 
mostly to $WW$ and $ZZ$. The capability of a linear $e^+e^-$ collider
with $\sqrt{s} = 500$ GeV
in this case has been analyzed in Ref.~\cite{fermi}.
More striking than the Higgs physics would be the non-standard phenomena
discussed above, associated with the KK modes.
A linear $e^+e^-$ collider operating at a center of mass energy of a few TeV
would probe the new physics at the electroweak, 
compactification, and string scales.

%
%
%
%




\begin{references}

\bibitem{Lee:1977yc}
B.~W.~Lee, C.~Quigg and H.~B.~Thacker,
``The Strength Of Weak Interactions At Very High-Energies And The Higgs Boson Mass,''
Phys.\ Rev.\ Lett.\ {\bf 38}, 883 (1977).

\bibitem{ewwg}
P.~Langacker,
``Physics implications of precision electroweak experiments,''
hep-ph/0102085; \\
J.~Erler,
``Fundamental parameters from precision tests,''
hep-ph/0102143.

\bibitem{Appelquist:1975tg}
T.~Appelquist and J.~Carazzone,
``Infrared Singularities And Massive Fields,''
Phys.\ Rev.\ D {\bf 11}, 2856 (1975).

\bibitem{Chivukula:2000px} R.~S.~Chivukula, C.~Holbling and
N.~Evans, 
``Limits on a composite Higgs boson,'' 
Phys.\ Rev.\ Lett.\  {\bf 85}, 511 (2000)
[hep-ph/0002022].

\bibitem{Antoniadis:1990ew} I.~Antoniadis, 
``A Possible New Dimension At A Few TeV,'' 
Phys.\ Lett.\  {\bf B246}, 377 (1990).

\bibitem{Dienes:1999vg} K.R.~Dienes, E.~Dudas and T.~Gherghetta, 
``Extra spacetime dimensions and unification,'' 
Phys.\ Lett.\ {\bf B436}, 55 (1998) 
[hep-ph/9803466].

\bibitem{Dobrescu:1999dg}
B.~A.~Dobrescu,
``Electroweak symmetry breaking as a consequence of compact dimensions,''
Phys.\ Lett.\ {\bf B461}, 99 (1999)
[hep-ph/9812349].

\bibitem{Cheng:2000bg}
H.~Cheng, B.~A.~Dobrescu and C.~T.~Hill,
``Electroweak symmetry breaking and extra dimensions,''
Nucl.\ Phys.\ {\bf B589}, 249 (2000)
[hep-ph/9912343].

\bibitem{Arkani-Hamed:2000hv}
N.~Arkani-Hamed, H.~Cheng, B.~A.~Dobrescu and L.~J.~Hall,
``Self-breaking of the standard model gauge symmetry,''
Phys.\ Rev.\ D {\bf 62}, 096006 (2000)
[hep-ph/0006238].

\bibitem{Luty:2000fj}
M.~A.~Luty, J.~Terning and A.~K.~Grant,
``Electroweak symmetry breaking by strong supersymmetric dynamics at the  TeV scale,''
hep-ph/0006224.

\bibitem{godbole}
R.~M.~Godbole and N. Arkani-Hamed, 
plenary talks at the Linear Collider Workshop 2000,
http://www-lc.fnal.gov/lcws2000/

\bibitem{Popovic:2000dx}
M.~B.~Popovic and E.~H.~Simmons,
``Weak-singlet fermions: Models and constraints,''
Phys.\ Rev.\ {\bf D 62}, 035002 (2000)
[hep-ph/0001302].

\bibitem{Dobrescu:1998nm}
B.~A.~Dobrescu and C.~T.~Hill,
``Electroweak symmetry breaking via top condensation seesaw,''
Phys.\ Rev.\ Lett.\ {\bf 81}, 2634 (1998)
[hep-ph/9712319].

\bibitem{Chivukula:1999wd}
R.~S.~Chivukula, B.~A.~Dobrescu, H.~Georgi and C.~T.~Hill,
``Top quark seesaw theory of electroweak symmetry breaking,''
Phys.\ Rev.\ D {\bf 59}, 075003 (1999)
[hep-ph/9809470].

\bibitem{private} R.~S.~Chivukula, private communication.

\bibitem{Collins:2000rz}
H.~Collins, A.~K.~Grant and H.~Georgi,
``The phenomenology of a top quark seesaw model,''
Phys.\ Rev.\ D {\bf 61}, 055002 (2000)
[hep-ph/9908330].

\bibitem{Dobrescu:2001gv}
B.~A.~Dobrescu,
``Minimal composite Higgs model with light bosons,''
Phys.\ Rev.\ D {\bf 63}, 015004 (2001)
[hep-ph/9908391].

\bibitem{Dobrescu:2000jt}
B.~A.~Dobrescu, G.~Landsberg and K.~T.~Matchev,
``Higgs boson decays to CP-odd scalars at the Tevatron and beyond,''
hep-ph/0005308.


\bibitem{Popovic:2001cj}
M.~B.~Popovic,
``Third generation seesaw mixing with new vector-like weak-doublet  quarks,''
hep-ph/0101123.


\bibitem{Hill:1991at}
C.~T.~Hill,
``Topcolor: Top quark condensation in a gauge extension of the standard model,''
Phys.\ Lett.\ {\bf B266}, 419 (1991).

\bibitem{Burdman:1999vw}
G.~Burdman and N.~Evans,
``Flavour universal dynamical electroweak symmetry breaking,''
Phys.\ Rev.\ D {\bf 59}, 115005 (1999)
[hep-ph/9811357].

\bibitem{Georgi:2001wt}
H.~Georgi and A.~K.~Grant,
``A topcolor jungle gym,''
Phys.\ Rev.\ D {\bf 63}, 015001 (2001)
[hep-ph/0006050].

\bibitem{He:2000vp}
H.~He, T.~Tait and C.~P.~Yuan,
``New topflavor models with seesaw mechanism,''
Phys.\ Rev.\ D {\bf 62}, 011702 (2000)
[hep-ph/9911266].

\bibitem{Aranda:2000vk}
A.~Aranda and C.~D.~Carone,
``Bounds on bosonic topcolor,''
Phys.\ Lett.\ {\bf B488}, 351 (2000)
[hep-ph/0007020].

\bibitem{Appelquist:2000nn}
T.~Appelquist, H.~Cheng and B.~A.~Dobrescu,
``Bounds on universal extra dimensions,''
hep-ph/0012100.

\bibitem{fermi} P. F.  Derwent, {\it et al},
``Linear Collider Physics'', report FERMILAB-FN-701, 
January 2001.



\end{references}
\end{document}